\documentclass[aps,prl,twocolumn,superscriptaddress,showpacs,floatfix]{revtex4}
\usepackage[cp866]{inputenc}
\usepackage[english]{babel}
\usepackage{bm}
\usepackage{times}
\usepackage{graphicx}
\usepackage{epsfig}
\usepackage[dvips,unicode,colorlinks,linkcolor=blue,citecolor=blue,urlcolor=blue]{hyperref}

\begin{document}

\title{Reduced Thermal Conductivity of Nanowires and Nanoribbons with Dynamically Rough Surfaces
and the ``Problem of One-Dimensional Heat Conductors''}
\author{Yuriy A. Kosevich}
\email[]{yukosevich@gmail.com}
\affiliation{Semenov Institute of Chemical Physics, Russian Academy of Sciences,
ul. Kosygina 4, 119991 Moscow, Russia}
\author{Alexander V. Savin}
\email[]{asavin@center.chph.ras.ru}
\affiliation{Semenov Institute of Chemical Physics, Russian Academy of Sciences,
ul. Kosygina 4, 119991 Moscow, Russia}


\begin{abstract}
We present analytical model and molecular dynamics simulations of phonon heat transport
in nanowires and nanoribbons with anharmonic lattices and dynamically rough surfaces
and edges. In agreement with recent experiments on heat transport in single-crystalline
silicon nanowires with rough surfaces, our model and simulations predict finite
and length-independent phonon thermal conductivity in such
quasi-one-dimensional systems,  in contrast to anomalous phonon thermal conductivity of
corresponding momentum-conserving
systems with atomically smooth surfaces,
divergent with the system length. Within our model, the main cause of thermal
conductivity reduction is momentum-nonconserving scattering of longitudinal acoustic
phonons by anharmonic side phonon leads in quasi-one-dimensional phonon waveguide
with dynamically rough surface or edge layers.
\end{abstract}


\maketitle

Thermal conductivity (TC) of low-dimensional nanostructures has recently attracted
much interest in connection with the search for technologically feasible materials
for thermoelectric applications \cite{dresselhaus07,tritt06}. High-efficiency
thermoelectric materials are important for power-generation devices, which are designed
to convert waste heat into electrical energy, and therefore they are expected to play
an increasingly important role in meeting the energy challenge of the future.
The reduction of the TC without significant reduction of the electric conductivity
is the most important strategy for enhancing the thermoelectric figure of merit, the
parameter which measures the material potential for thermoelectric applications
\cite{dresselhaus07,tritt06}. Recent experiments have shown that TC of silicon
nanowires (NWs) can be dramatically reduced by surface roughness \cite{hochbaum,boukai}.
Furthermore, the phonon contribution to TC of a single-crystalline Si NW with diameter
$D\sim 50$ nm was shown to approach the limit of amorphous Si and therefore a
single-crystalline Si NW with such small diameter behaves like a "phonon glass"
 \ \cite{hochbaum}. This observation cannot be explained by the existing theories,
although the important role of surface roughness in phonon scattering in NWs has been
emphasized  \cite{santamore01,li03,murphy,donadio}. Molecular dynamics (MD)
modeling of diamond nanorods with functionalized surfaces
also shows significant TC reduction in such low-dimensional systems  \cite{padgett}.

On the other hand, quasi-ballistic long wavelength acoustic phonons in disordered
harmonic or ideal anharmonic one-dimensional (1D) systems give rise to TC coefficient, divergent
with the system length, see, e.g.,
\cite{pomeranchuk,ziman,maynard,lepri,kambili,narayan,mingo05}. As it was shown
for the first time by Peierls, only the  {\it momentum-nonconserving} scattering of phonons
can limit TC of the system \cite{peierls}.  In this Letter we present
analytical model and MD simulations of the TC of NWs and nanoribbons (NRs) with
dynamically rough surfaces and edges which show dramatic  decrease of TC in such
partially-disordered systems, in contrast to TC of corresponding ideal systems
(with atomically smooth surfaces and edges). The analytical model describes 1D lattice
of coupled oscillators in which each oscillator is weakly coupled to 
weakly-anharmonic side
oscillatory chain, or {\it side phonon lead} (SPL),
which models {\it dynamical surface roughness} of the quasi-one-dimensional phonon
waveguide (Q1DPW), see Ref.  \cite{kos1}. Within this model, the main cause of TC
reduction is momentum-nonconserving scattering of longitudinal acoustic phonons
by SPLs in dynamically rough surface or edge layers of the Q1DPW.
Strong scattering of coherent longitudinal acoustic phonons in the Q1DPW is a result of
energy losses caused by the excitation of acoustic phonons in SPLs, which propagate
only {\it outwards} and do not return back to the Q1DPW in the form of coherent waves
\cite{kos1}. This occurs when the length of a SPL exceeds acoustic phonon mean
free path (localization length) $l_{ph}^{(spl)}$ due to anharmonic
scattering in the lead (which gives $l_{ph}^{(spl)}$$\propto$$\omega^{-2}$).
The effective length of
the atomic or molecular
SPL
increases by random intersections with other SPLs in the rough surface or edge layer.

The proposed phonon scattering mechanism gives finite and frequency-independent mean
free path $l_{ph}$ of the propagating along Q1DPW longitudinal acoustic phonons
in a wide frequency range, including low frequencies \cite{kos1}. The value of phonon
mean free path due to the dynamical-roughness-induced momentum-nonconserving scattering
is given by the lattice spacing in the Q1DPW $a$ and coupling  parameter  between a SPL
and Q1DPW $g$: $l_{ph}$$\approx$$a/g$. Coupling parameter $g$ is determined by the ratio
of the cross-section area of the rough layer ($\sim$$\pi D\delta$)
with the average
root-mean-square roughness height $\delta$ and cross-section area of NW with diameter
$D$: $g$$\sim$$\delta/D$ \cite{kos1}. For atomically-thick rough surface layer with
$\delta$$>$a, acoustic phonon mean free path $l_{ph}$$\sim$$D(a/\delta)$
can become {\it shorter} than NW diameter \cite{kos1}. Therefore this channel of surface phonon
scattering substantially reduces mean free path of \ ``long longitudinal waves'' which
give dominant  contribution to phonon TC of quasi-1D systems,
see Refs. \cite{pomeranchuk,ziman,mingo05}. This in turn causes a substantial reduction
of phonon TC in NWs and NRs with dynamically rough surfaces or edges, which explains
the  significant TC reduction, down to the level of TC of amorphous Si, observed in
single-crystalline Si NWs with rough surfaces \cite{hochbaum}. The predicted
reduction factor $a/\delta$ is quantitatively consistent with the observed reduction
of TC of Si NWs with rough surface layers with controlled thicknesses
$\delta$$>$$a$ \cite{hochbaum} with respect to TC of Si NWs with the same diameters and
smooth surfaces (with $\delta$$\sim$$a$) \cite{kos1}.

This mechanism of momentum-nonconserving acoustic phonon scattering gives {\it finite}
and {\it length-independent} coefficient of TC of Q1DPWs with atomically rough surfaces
or edges with $\delta$$>$$a$, in contrast to the divergent with the system length coefficient of TC 
of anharmonic Q1DPWs with {\it atomically smooth}  surfaces
or edges with $\delta$$\ll$$a$,
when the  momentum-nonconserving channel of acoustic phonon scattering is closed and
mean free path of long-wave phonons substantially increases, $l_{ph}$$\propto$$\omega^{-2}$, and becomes $l_{ph}$$\gg$$D$.
The considered mechanism of acoustic phonon scattering
by vibration of atomic or molecular chains in dynamically rough
surfaces and edges of NWs and NRs
is qualitatively different from usually considered acoustic wave scattering by
static (geometric) roughness on stress-free surfaces of solids \cite{santamore01,kambili}.

{\bf Analytical model of phonon transport in quasi-1D phonon waveguides with
side phonon leads}.
We consider 1D lattice with period $a$ of coupled oscillators,
with coupling constant $c$ and mass $m$,
in which each oscillator is weakly coupled to SPL (``lateral''
atomic chain with weakly-anharmonic coupling), with (linear) interatomic coupling $c_l$=$gc$$\ll$$c$ and
atomic mass $m_l$=$gm$$\ll$$m$. Distribution of such
SPLs along the Q1DPW models its dynamical surface roughness \cite{kos1}. The most
important assumption of this model is that acoustic phonon, propagating along the
Q1DPW, excites acoustic waves in SPLs which propagate along the leads only outwards
the Q1DPW and do not return back to the waveguide in the form of coherent waves.
With this assumption, we obtain 
dispersion equation for the
dimensionless complex wave number $k_{\parallel}a$ of the damped acoustic phonons,
propagating along the Q1DPW \cite{kos1}:
    \begin{equation}
    k_{\parallel}^2a^2=\omega^2\frac{m}{c}+i\omega g \sqrt{\frac{m}{c}}.
    \end{equation}
This equation predicts that phonons with frequencies
$\omega\gg\omega^\ast\equiv g\sqrt{c/m}$ propagate quasi-ballistically, when
    \begin{equation}
    k_{\parallel}a=\omega\sqrt{\frac{m}{c}}+\frac{i}{2}g,
    \end{equation}
while low-frequency phonons, with $\omega$$ \ll$$\omega^\ast$, propagate diffusively,
when $\omega$$=$$-iD_{ph}$$ k_{\parallel}^2$, with the diffusion coefficient
$D_{ph}$=$a^2\sqrt{c/m}/g$=$V_{ph}l_{ph}$,
where $V_{ph}$=$a$$\sqrt{c/m}$ and $l_{ph}$=$a/g$ are velocity and energy mean free path
of long-wave acoustic phonons, cf. Eq. (2). Both quasi-ballistic and diffusive
acoustic phonons contribute to phonon TC coefficient $\kappa_{ph}$ of Q1DPW
as $\kappa_{ph}$$\sim$$ C_{ph}$$D_{ph}$$\sim$$ C_{ph}$$V_{ph}$$a/g$,
where  $C_{ph}$ is the contribution of longitudinal phonons to the specific heat of NW or NR.
This predicts, in agreement with the observations \cite{hochbaum,boukai} and our MD
simulations,
the significant reduction of phonon mean free path
and TC caused by phonon scattering by dynamical surface roughness in
NWs and NRs
(with $g$$\sim$$\delta$$/D$ and  $\delta$$\gg$$a$).

{\bf MD simulations of TC of 2D nanoribbons}.
We consider the atomic structure consisting of $M$ parallel chains
of atoms placed in one plane.
We consider the Hamiltonian of the
scalar model of the ribbon lattice in which only the longitudinal
displacements of atoms are taken into account and
displacement of the $(m,n)$th atom from its equilibrium position. The Hamiltonian,
which accounts for interaction between only the nearest-neighbor atoms (with unit mass):
    \begin{eqnarray}
    H&=&\sum_{m=1}^M\sum_{n=1}^N\frac12\dot{u}_{m,n}^2+\sum_{m=1}^M\sum_{n=1}^{N-1}
    V_{mn}(u_{m,n+1}-u_{m,n})\nonumber \\
    &&+\sum_{m=1}^{M-1}\sum_{n=1}^NU_{mn}(u_{m+1,n}-u_{m,n}),
    \label{f1}
    \end{eqnarray}
where $N$ is number of atoms in each chain, $V_{mn}$ and $U_{mn}$ are potentials
of the intra- and interchain interaction between $(m,n)$ and $(m,n+1)$ atoms, and
between $(m,n)$ and $(m+1,n)$ atoms, respectively.
For the purpose of calculating the heat flux along the NR, we can define 
the corresponding total energy of the $n$th cross-section
$e_n$ and local energy flux $j_n$ in the ribbon, which satisfy the
continuity condition $\dot{e}_n$=$j_{n-1}-j_{n}$, see, e.g.,  Ref. \cite{sav1}.

\begin{figure}[t]
\includegraphics[angle=0, width=1\linewidth]{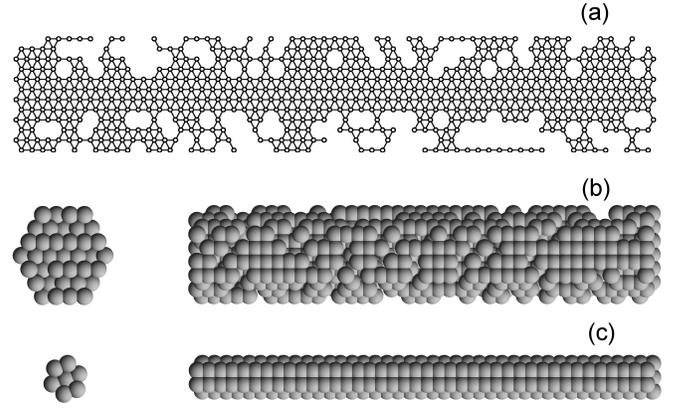}
\caption{
(a) Model of 2D atomic ribbon with rough edges which consists of $M$=12 chains
with ideal core of $M_0$=4 chains and two rough edges of $M_1$=4 chains with $p$=0.3
randomly deleted atoms.
(b) Model of 3D nanowire with rough surface layers of 2 atom thickness.
(c) Ideal 3D core. }
\label{fig1}
\end{figure}

We consider a NR built of $M=M_0+2M_1$ chains. To model two
dynamically rough edges with widths $M_1$,  we randomly
delete some atoms in the chains
$m=1,...,M_1$ and $m=M_1+M_0+1,...,M_1+M_0+M_1$. Let $0\le p\le 1$ be the
probability of atom removal. In result of the random atom removal from the edge layers,
some atoms in the edges will become completely isolated and should be deleted as well.
Model NR with two rough edges with $p=0.3$ randomly deleted atoms
is shown in Fig. \ref{fig1}(a).

Assuming for the certainty that
$U(\rho)=\frac12V(\rho)$, where $V(\rho)\equiv V_{mn}(\rho)$ and $U(\rho)\equiv U_{mn}(\rho)$,
we consider  $V(\rho)=\rho^2 /2$, $V(\rho)=\rho^2/2+\rho^4/4$,
$V(\rho)=\exp(-\rho)+\rho -1$, $V(\rho)=[\exp(-\rho)-1]^2/2$ and
$V(\rho)=1-\cos(\rho)$ as, respectively,
the harmonic, Fermi-Past-Ulam (FPU), Toda, Morse and rotational potentials.
It is worth mentioning that the ribbons with the harmonic, FPU, Toda or Morse interatomic
potentials and ideal atomically smooth edges have infinite coefficients of TC in the
limit of $N\rightarrow\infty$. In purely harmonic systems, acoustic phonons do not
interact and there is no energy scattering in phonon thermal transport. Infinite
TC coefficients in ribbons with the anharmonic FPU, Toda or Morse  interatomic potentials
are related with the quasi-ballistic transport of the long-wave acoustic phonons
with long mean free paths $l_{ph}\propto \omega^{-2}$. Ideal ribbons with the
rotational potential have finite TC in the limit of $N\rightarrow\infty$ and finite temperatures \cite{sav2,politi}.

The ribbon TC is found in two independent ways. The first method relies on direct
modeling of heat transport along the ribbon. For this purpose, we consider a ribbon
with total length $N+2N_0$ ($N_0=40$ is the length of ribbon ends), which ends
are placed in Langevin thermostats with different temperatures
of the left, $T_+=1.1T$, and right,$T_-=0.9T$, ends, where $T$ is average temperature).
In the middle part of the ribbon with length $N$ the stationary heat flux $J_n=J$
and linear temperature gradient of ribbon temperature $T_n$ are established (where $J_n$
is determined by the time-average of the local heat flux $\left\langle j_n\right\rangle$
in the stationary conditions).  Thus the TC coefficient $\kappa(N)$ of the finite-length
ribbon can be reliably determined as
    \begin{equation}
    \kappa(N)=\frac{J(N-1)}{(T_{N_0+1}-T_{N_0+N})(M_0+2M_1q)},
    \label{f8}
    \end{equation}
where $M_0$ is the width of central ideal strip,
$2M_1 q$ is total width of rough edges with 
filling fraction $q$$\equiv$$1-p<1$.
TC coefficient $\kappa$ is determined from $\kappa(N)$ as the limit of $N\rightarrow\infty$.
The ribbon temperature profile $T_n$ depends on a particular realization of the edge
roughness and therefore it is necessary to perform averaging over its independent
realizations.
In our simulations,
we averaged over 120 independent realizations of the edge roughness.

\begin{figure}[tp]
\includegraphics[angle=0, width=1\linewidth]{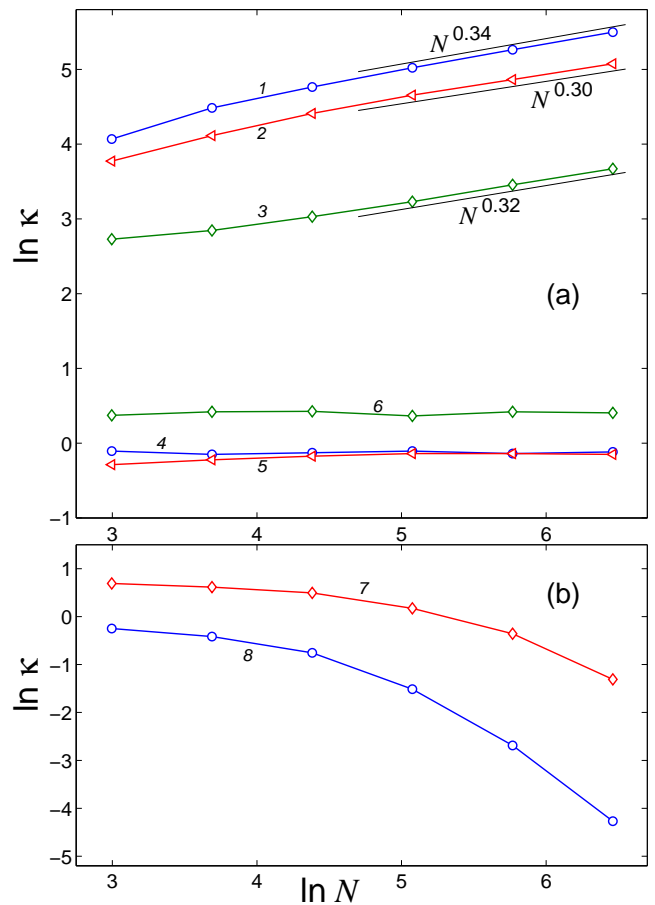}
\caption{
(a) Dependence of TC coefficient $\kappa$ on length $N$ of 2D ribbons with smooth
and rough edges for the Toda (curves 1 and 4, temperature $T=1$),
Morse (curves 2 and 5, temperature $T=0.1$) and FPU (curves 3 and 6, $T$=10)
interatomic potentials.
(b) Dependence of TC coefficient $\kappa$ on length $N$ of 2D ribbons with purely 
harmonic lattices  and rough edge widths $M_1$=1 and $M_1$=4 (curves 7 and 8)
with $p$=0.3 randomly deleted atoms. Ideal core width is
$M_0$=2, temperature $T$=1.
 }
\label{fig2}
\end{figure}

We modeled heat transport in rough-edge NRs with central part length of
$N$=$20$,$40$,$80$,$160$,$320$,$640$. Dependence of TC coefficients on the length of
NR with rough edges, with $p=0.3$ randomly deleted atoms, is shown
in Fig. \ref{fig2}(a) for different interatomic potentials. As one can see in
this figure, the rough-edge ribbon has finite coefficient of TC in the limit
of infinite length for all the considered nonlinear interatomic potentials.
In contrast to that, TC coefficient $\kappa(N)$ of the rough-edge ribbon with purely harmonic
lattice continuosly decreases for $N$$\rightarrow$$\infty$, see Fig. \ref{fig2}(b),
and the limiting value of TC coefficient of such NR is zero,  $\kappa(\infty)$=$0$.
Figure 2(b) also shows that for the given width of the ideal ribbon core $M_0$,
TC coefficient of the NR decreases with the increase of the rough edges width $M_1$.
Therefore the infinite-length ribbon or wire with purely harmonic lattice and
dynamically rough edges
can be considered as an ideal {\it thermal insulator}.

We relate the reason for such strong difference between TC coefficients of
rough-edge NRs with harmonic and anharmonic interatomic potentials with the
properties of SPLs in corresponding systems. In the anharmonic systems,
long acoustic waves, excited in SPLs, do not return back and the effective
``internal radiative losses'' result in finite phonon mean free path and
phonon TC in the Q1DPW, see Eqs. (1) and (2). In purely harmonic systems,  the long
acoustic waves, excited in SPLs, return back into the Q1DPW and strongly suppress the 
transmission of low-frequency acoustic phonons through the Q1DPW because of destructive
interference with the phonons propagating along the waveguide core
(the {\it phonon Fano resonance}, see Ref. \cite{kos1} for a brief review). This
produces the effective stop band (or Anderson-Fano localization) for low-frequency
acoustic phonons in the Q1DPW with harmonic lattice and dynamically rough
edges \cite{kos1}, and, correspondingly, the decrease of TC coefficient of the NR
in the limit of $N$$\rightarrow$$\infty$. Indeed, we observe only the localized
vibration eigenstates in harmonic phonon spectrum of such system (not shown).

In the approach
based on Green-Kubo method, TC is defined as integral of the autocorrelation of heat fluxes:
\begin{equation}
\kappa(N)=\lim_{t\rightarrow\infty}
\frac{1}{NT^2(M_0+2M_1q)}\int\limits_0^t\langle J(\tau)J(\tau-t)\rangle d\tau,
\label{f9}
\end{equation}
where $J(t)$=$\sum_{n}j_n(t)$ is the average heat flux in the NR.
Values of $\kappa(N)$ obtained with the use of two different approaches, Eqs.(\ref{f8})
and (\ref{f9}), coincide with good accuracy.

We also perform MD simulations of the temperature dependence of TC
of rough-edge NRs with the FPU, Toda and rotational interatomic potentials (not shown).
All the obtained dependencies confirm the finite coefficients of TC of the NRs in the low temperature
limit $T\rightarrow 0$ when the coefficients of classical TC of corresponding NRs with atomically smooth edges diverge.
\begin{figure}[tp]
\includegraphics[angle=0, width=1\linewidth]{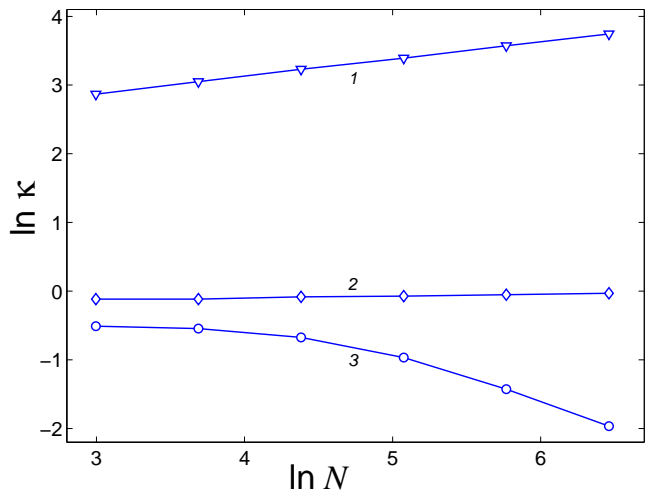}
\caption{
Dependence of TC coefficient of nanowires with ideal (curve 1) and rough
(curves 2 and 3) surfaces on length $N$, for the Toda (curves 1 and 2)
and harmonic (curve 3) potentials.
}
\label{fig3}
\end{figure}

{\bf MD simulations of TC of 3D nanowires}. We consider 3D model system made of
$M\times K$ parallel molecular chains as it is shown in Figs. \ref{fig1}(b)
and \ref{fig1}(c).
The Hamiltonian of the system we take as a 3D generalization of the Hamiltonian (3)
of 2D ribbon scalar model.  In Fig. \ref{fig3} we show the length dependence of nanowire TC
coefficient for the Toda and harmonic interatomic potentials. It demonstrates that
similar to the case of 2D nanoribbon, the dynamical roughness of the surface layer
changes the increasing (divergent)
with the length TC of 3D nanowire to the finite or decreasing with the
nanowire length TC of 3D nanowire with, respectively, the anharmonic or
harmonic interatomic potential.

In summary, we present analytical model and molecular dynamics simulations of
phonon heat transport in nanowires and nanoribbons with anharmonic lattices and
dynamically rough surfaces and edges. In agreement with recent experiments on
heat transport in single-crystalline silicon nanowires with rough surfaces, our
model and simulations predict finite and length-independent phonon thermal
conductivity in such quasi-one-dimensional systems, in contrast
to anomalous phonon thermal conductivity of corresponding momentum-conserving
systems with atomically
smooth surfaces, divergent with the system length. We also present the
thermal-insulator-like heat  transport in
long-length nanowires and nanoribbons with purely harmonic lattices and
dynamically rough surfaces, caused by  the Anderson-Fano phonon localization.

Yu.A.K. thanks A. Cantarero, L.B. Dubovskii, Yu.M. Kagan and O.V. Rudenko for useful discussions.
The authors thank RFBR for the support through Grant No. 08-03-00420 and JSC RAS for
using computer facilities.

\end{document}